# Nonlinearity Modulation of Auto-oscillations in Three-terminal Magnetic Tunnel Junctions

Zixi Wang [1], Wenlong Cai [1,2,*], Ao Du [1], Zanhong Chen [1], Lei Zhou [1], Shiyang Lu [1], Kewen Shi [1,2,*] and Weisheng Zhao [1,2,*]

[1]*Fert Beijing Institute, School of Integrated Circuit Science and Engineering, Beihang University, Beijing 100191, China*
[2]*National Key Laboratory of Spintronics, Hangzhou International Innovation Institute, Beihang University, Hangzhou 311115, China*

Spin torque nano-oscillators (STNOs) hold encouraging promise for nanoscale microwave generators, modulators, and new types of intelligent computing. The nonlinearity, describing the current-induced tunability of oscillating frequency, is a distinctive feature of STNOs, which plays important roles in efficient manipulation of microwave frequencies, rapid spectrum analysis, and the design of neuromorphic devices. However, experimental research on its efficient modulation remains limited. Here, we comprehensively studied the impact of several factors on nonlinearity in nanoscale three-terminal MTJ-STNOs, including the external magnetic field, the thickness of CoFeB free layer, and the combination of spin-transfer torque (STT) and spin-orbit torque (SOT). Among these factors, nonlinearity can be significantly tuned by the direction of magnetic field as well as the thickness of CoFeB free layer. Notably, it reaches zero in 1.1 nm CoFeB, where the oscillation frequency is not affected by the drive current. Such property provides a more intrinsic and robust approach to achieve zero nonlinearity in STNOs, which is advantageous for high-quality microwave generators. More importantly, we found that nonlinearity can also be electrically modulated by both STT and SOT currents, and develop a refined model that accounts for the additional contribution of the SOT current to explain the mechanism. This electrical approach is more convenient, energy-efficient, and well-suited for miniaturization. Our findings offer a comprehensive understanding and open up a new dimension for the current tunability of nonlinearity in MTJ-STNOs, benefiting further optimization in nanoscale STNO-based microwave generators and neuromorphic computing devices.

**spintronics, spin torque nano-oscillator, magnetic tunnel junction, nonlinear auto-oscillator theory, spin orbit torque**

## 1 Introduction

Spin torque nano-oscillators (STNOs) [1–3] based on magnetic tunnel junction (MTJ) [4] can be driven into auto oscillation with output microwave signals by the current-induced spin-transfer torque (STT). With the advantages of high current tunability, size in nanoscale, and wide output bandwidth, STNO is one of the reliable candidates for nanoscale microwave generators [5,6], detectors [7–9], spectral analyzers [10], and amplifiers [11]. Besides, utilizing the intrinsic nonlinearity of the oscillating frequency, MTJs in nanoscale can also serve as spintronic synapses [12–14] and neurons [15–17], providing potential applications in neural networks (NNs) [18] and other neuromorphic computing scenarios [19–22]. On the other hand, spin Hall nano-oscillators (SHNOs) [23,24] driven by spin-orbit torque (SOT) in the heavy metal layer possess higher energy efficiency than STT and potential for mutual synchronization [25,26]. Nevertheless, the low power emission caused by the extremely low magnetoresistance (MR) in SHNOs remains the primary challenge of further use. A three-terminal STNO, however, combines the advantages of large emission power (large MR) and the high energy efficiency of SOT [27,28], Besides, the combination of STT and SOT enables a larger tunable range for oscillators, possessing great potential for neuromorphic computing [29].

The modulation of oscillation frequency by current plays a crucial role in various STNO-based applications. In order to characterize this behavior, nonlinear auto-oscillation theory [30–32] introduces a nonlinear frequency shift modulated by the dc current. The nonlinear factor $N$ describes the current tunability of the STNO's output frequency. Based on the theory, the nonlinearity $N$ can be modified by the magnitude and direction of the external magnetic field $H_{ex}$ [33–35], as well as the effective magnetization $M_{eff}$ and the magnetic anisotropy of the free layer (FL) [32,36]. In radiofrequency (RF) spintronic devices, such as microwave generators and modulators [5], nonlinearity determines the large output bandwidth tuned by dc current, and the linewidth can also be reduced significantly when $N$ approaches zero [36,37]. Besides, as synapses and neurons in neuromorphic computing architectures, nonlinearity of STNOs was utilized to emulate certain behaviors of the networks, such as the resonance frequency and synaptic weight [18]. Previous experimental studies of nonlinearity mainly focus on the magnitude and direction of $H_{ex}$ [38], but the strict constraint on $H_{ex}$ highly limits the output frequency. By using He$^+$ irradiation to modify the interfaces of the FL, the nonlinearity $N$ of STNOs was successfully modulated with different $M_{eff}$ [36]. Besides, the nonlinearity in SHNO was also investigated to explored its potential in neuromorphic computing applications [25,39]. However, experimental research on the efficient modulation of nonlinearity in three-terminal STNOs remains limited, especially by the combination of STT and SOT effects.

In this work, we comprehensively clarified the impact of several factors on nonlinearity in three-terminal MTJ-STNOs, including the magnitude and direction of $H_{ex}$, the thickness of CoFeB FL, and the combination of STT and SOT currents. As a result, nonlinearity reaches zero in 1.1 nm CoFeB, providing a robust and intrinsic approach that is advantageous for high-quality microwave generators. Additionally, we proved that nonlinearity can be modulated by the both STT and SOT currents, and rectified the nonlinearity theory by considering both currents to explain this phenomenon. This electrical approach is more convenient, energy-efficient, and well-suited for miniaturization. Those modulating methods provide a comprehensive understanding of the modulation of nonlinearity in MTJ-STNOs, and will pave the way for further optimization in STNO-based applications.

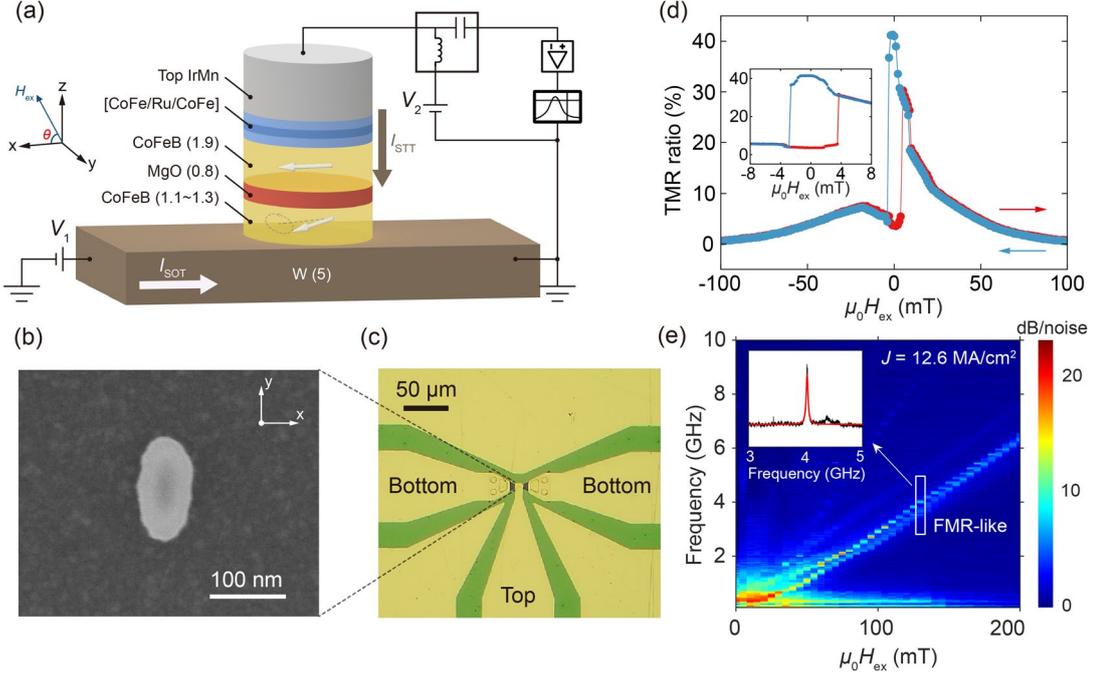

**Figure 1** STNO's geometry and basic magnetic properties. (a) Schematic and experimental setup of the three-terminal STNO. The (b) scanning electron microscope (SEM) image and (c) optical microscopy image of the three-terminal device. The shape anisotropy and magnetic anisotropy of the MTJ are along the y and x direction, respectively. (d) TMR ratio as a function of the external magnetic field $H_{ex}$ sweeping along x direction. The inset shows the minor loops of TMR ratio. (e) PSD as a function of $H_{ex}$ in 1.3 nm CoFeB. The inset shows a single PSD spectrum (black curve) when $\mu_0 H_{ex} = 130$ mT and $\theta = 30°$, and the red curve represents the fitted Lorentzian function.

## 2 Materials and method

### 2.1 Sample fabrication

Figure 1(a) illustrates the schematic of the proposed three-terminal STNO device. The film stacks were deposited on thermally oxidized Si/SiO$_2$ substrate, composed of substrate/W (5.0)/Co$_{20}$Fe$_{60}$B$_{20}$ (1.1~1.3)/MgO (0.8)/ Co$_{20}$Fe$_{60}$B$_{20}$ (1.9)/CoFe$_{30}$ (0.5)/Ru (0.8)/CoFe (2.0)/ Ir$_{20}$Mn$_{80}$ (7.5)/Ru (2.0)/Ta (3.0)/Ru (10.0) (numbers in the parentheses denote thickness in nanometers). The CoFeB/CoFe composite serves as the reference layer (RL), while the CoFeB below represents the FL. The thickness of the FL is designed to be 1.1 nm and 1.3 nm, respectively, aiming to achieve different magnetic anisotropy and demagnetizing field $H_k$. Subsequently, the deposited stacks were annealed for 1 hour at 300 °C under a magnetic field of 1 T (x direction), and cooled to room temperature later. Thereafter, the full stack was patterned into the shape of bottom SOT channel. The stack on the W bottom electrode was then patterned into nanoscale circular (80 nm) and elliptical (80 nm × 120 nm) shapes by electron beam lithography and ion beam etching Figure 1(b). Finally, the top of the MTJ pillars were connected with the top electrode using lift-off process, and result in three-terminal STNOs (Figure 1(c)). Figure 1(d) shows the resistance loop as a function of the external magnetic field, and a TMR ratio of 41.5% was measured in the STNO. Figure 1(e) illustrates the power spectral density (PSD) vs. $H_{ex}$ at a fixed STT current, and the output signal corresponds to the FMR-like mode. All PSD spectra fit well with the Lorentzian function (inset of Figure 1(e)), from which we can extract the frequency, power and linewidth of a specific signal.

## 2.2 Device characterization

All electrical properties of the STNO devices were characterized on a custom-built probe station presented in Figure 1(a). The amplified microwave signal was detected by a spectrum analyzer. The DC current was applied separately to the MTJ and the SOT channel via the dual channels of the Keysight B2912A current source. The RF signal generated by the oscillation of the STNO is amplified by an active amplifier, and detected through Keysight N9020B MXA signal analyzer. A bias-T is used to separate the DC and RF signals on the STNO. The external magnetic field is applied using a magnet coil that can rotate within the x-z plane. The angle between external field $H_{ex}$ and the x-axis is denoted as $\theta$. All measurements were performed at atmosphere and room temperature.

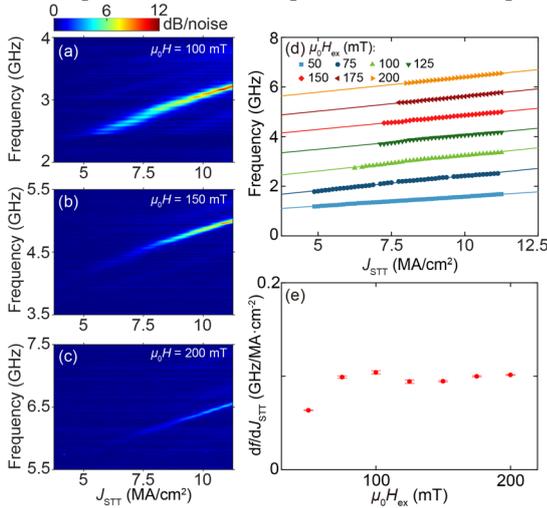

**Figure 2** Modulation of STNO's nonlinearity $N$ by the magnitude of the external magnetic field in 1.3 nm FL: (a)-(c) PSD vs. STT current density under different $H_{ex}$, the angle $\theta$ is fixed at 0° (in-plane). (d) Current tunability of the peak frequency under different $H_{ex}$. The data points are experimental results while the solid lines represent the fitting curves. (e) The slope $df/dJ_{STT}$ as a function of the magnetic field.

## 3 Results and discussion

### 3.1 Modulation of nonlinearity by the external magnetic field

With the proposed STNO devices, we performed detailed measurements to examine the variation of nonlinearity with several factors, including the magnitude and angle of $H_{ex}$, the thickness of the FL, and the combination of STT and SOT current. Figure 2(a)-(c) presents the relationship between the PSD and the STT current density $J_{STT}$ under different in-plane $H_{ex}$ in a 1.3 nm CoFeB FL. The frequency exhibits a clear linear dependence on the current, and was fitted with linear function (Figure 2(d)). The slope $df/dJ_{STT}$ as a function of the magnitude of $H_{ex}$ is summarized in Figure 2(e). The fitted lines appear nearly parallel, and the slope also shows minimal variation (except the slightly smaller $df/dJ_{STT}$ observed at 50 mT) with changes in the magnetic field.

To comprehend the underlying mechanism of this phenomenon, we introduce the nonlinear auto-oscillator theory proposed by A. Slavin and V. Tiberkevich et al. [30,32], which provides an analytical model based on the universal model of an auto-oscillator. According to the theory, the current dependence of the auto-oscillation frequency $f$ generated from an STNO is expressed as:

$$f(J) = f_{FMR} + \frac{N}{2\pi}P, \quad P = |c|^2 = \frac{\zeta - 1}{\zeta + Q}, \quad (1)$$

where $J$ represents the current density of the driving current (typically the STT current), and $f_{FMR}$ is the FMR frequency of STNO. $P$ and $c$ denote the normalized power and amplitude of the stationary precession. $Q = 2\omega_M/\omega_0 - 1$ is the nonlinear damping coefficient, and $\zeta = J/J_{th}$ denotes the dimensionless supercriticality parameter, where $J_{th}$ represents the threshold current density of microwave generation. Thus, the nonlinearity $N$ can be derived from the $J - f$ curve:

$$\frac{df}{d\zeta} = J_{th}\frac{df}{dJ} = \frac{N}{2\pi}\frac{1+Q}{(\zeta + Q)^2}. \quad (2)$$

Therefore, the value of the nonlinearity factor $N$ is proportional to $df/dJ$, i.e. the current tunability of STNO. The frequency shows linear dependency on current density (Figure 2(d)), indicating the negligible variation of $\zeta$ compared with $Q$ in denominator. Assuming the external field is rather large and aligned with the magnetization of the FL, $N$ can be determined by [30]:

$$N = -\frac{\omega_H \omega_M \left(\omega_H + \frac{\omega_M}{4}\right)}{\omega_0 \left(\omega_H + \frac{\omega_M}{2}\right)}, \quad (3)$$

where

$$\begin{cases} \omega_H = \gamma H, \\ \omega_M = 4\pi\gamma M_{eff}, \\ \omega_0 = \gamma\sqrt{\omega_H(\omega_H + \omega_M)}. \end{cases} \quad (4)$$

$\gamma$ is the gyromagnetic ratio, and $\mu_0 M_{eff} = \mu_0 M_s - \mu_0 H_k$ is the effective magnetization. Based on eqs. (3) and (4), the magnetic field can directly modulate the nonlinearity $N$ by changing the magnitude of $\omega_H$. However, no significant changes are observed in $df/dJ_{STT}$ (i.e. the nonlinearity) in Figure 2(e). Such phenomenon could be attributed to the relatively small $M_{eff}$ of the FL or the factor $(1+Q)/(\zeta+Q)^2$ in eq. (2). Additionally, the range of $H_{ex}$ variation may not be large enough, leading to negligible changes of $df/dJ_{STT}$.

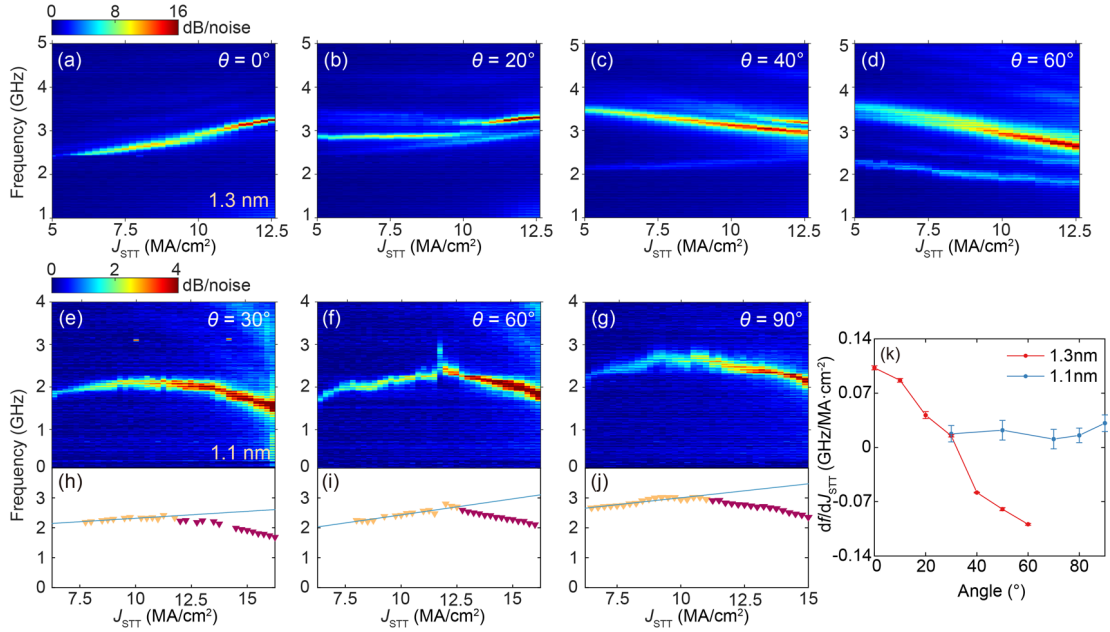

**Figure 3** The dependence of nonlinearity $N$ on the magnetic field angle $\theta$. PSD vs. STT current density under different angle $\theta$ of the magnetic field in (a)-(d) 1.3 nm CoFeB and (e)-(g) 1.1 nm CoFeB, respectively. The magnitude of the external field is fixed at $\mu_0 H_{ex} = 100$ mT. (h)-(j) The peak frequency as a function of $J_{STT}$ derived from the PSD in (e)-(g), respectively. The solid lines are linear fitting results in small current region (yellow dots), and the red dots are excluded due to the thermal effect. (k) $df/dJ_{STT}$ as a function of $\theta$ in samples with different FL thicknesses.

Figure 3(a)-(d) illustrates the auto-oscillation behavior at different angle of $H_{ex}$. Under a fixed magnitude of 100 mT, the slope $df/dJ_{STT}$ continuously changes with the angle $\theta$, distinctly indicating the nonlinearity $N$ shifts from positive ($\theta = 0°$) to negative ($\theta = 60°$). Mode hopping also appears at specific level of field and current. Here, we focus on the dominant mode and the extracted slope as a function of $\theta$ is plotted in Figure 3(k). As the angle $\theta$ increases, the precession axis of the FL magnetic moment, as well as the effective magnetic field $H_{eff}$, gradually tilts from in-plane to perpendicular direction. The projection of the $H_{ex}$ in the perpendicular direction increases, leading to a shift of $\omega_H$ in eqs. (3) and (4). The nonlinearity $N$ thus manifests a strong dependence on the angle of the $H_{ex}$.

### 3.2 Modulation of nonlinearity by the thickness of FL

To further investigate the impact of different FL thicknesses on nonlinearity $N$, we carried out similar measurements on STNO with 1.1 nm FL. As shown in the PSD spectrums in Figure 3(e)-(g), the oscillation frequency seemingly exhibits a quadratic dependence on the $J_{STT}$, rather than a linear dependence predicted by the nonlinear auto-oscillation theory [32]. This observation can be elucidated by the current-induced Joule heating, which leads to modifications of $H_k$ and $M_s$ of the FL [40] but is not taken into account in the theory. Due to the variations in the device fabrication process, thermal effects are pronounced in the 1.1 nm sample. Therefore, data points in the high current region (red dots in Figure 3(h)-(j)) were excluded to eliminate the influence of heating. Interestingly, compared with the oscillation behavior in 1.3 nm CoFeB, $df/dJ_{STT}$ of 1.1 nm CoFeB shows relatively small variation with the magnetic field angle $\theta$ (Figure 3(k)), indicating an extremely low nonlinearity $N$. We attribute this phenomenon to the small $M_{eff}$ of 1.1 nm CoFeB, caused by the rather high $H_k$ in a thinner film that compensates $M_{eff}$ to be around zero. Thus the $\omega_M$ in eqs. (3) and (4) become small enough, that the change of $H_{ex}$ has a minimal effect on nonlinearity $N$. As a result, it is demonstrated that, by modulating the thickness of CoFeB FL, nonlinearity $N$ can be continuously tuned, and reaches $N \to 0$ at 1.1 nm CoFeB FL. It provides us a more intrinsic and robust approach to attain zero nonlinearity in STNOs, not constrained by external factors such as direction and magnitude of the magnetic field. With such property, the output power of the STNO remains unaffected by the driving current. Moreover, small nonlinearity also has the significant advantage of reducing the output linewidth of the STNO, which is of great importance for commercializing microwave generators.

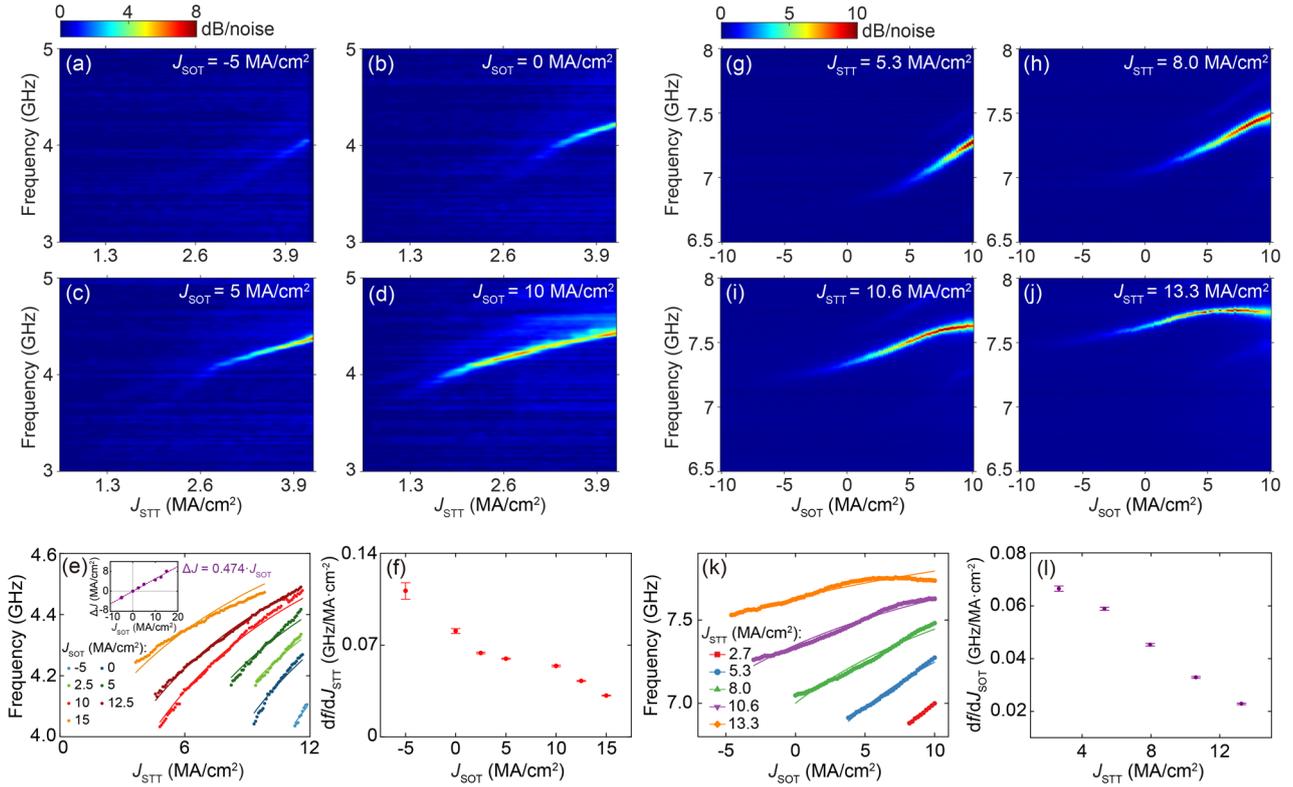

**Figure 4** Modulation of nonlinearity $N$ by the combination of the SOT and STT current in 1.3 nm CoFeB. (a)-(d) PSD vs. STT current density under fixed SOT currents and an in-plane magnetic field $\mu_0 H_{ex} = 140$ mT. (e) Oscillating frequency as a function of $J_{STT}$ under different $J_{SOT}$. Solid lines are the theoretical calculation from eq. (5). The inset illustrates the relationship between $J_{SOT}$ and the fitting parameter $\Delta J$, and the value of $\eta$ can be estimated through linear fitting (plotted as solid line). (f) The extracted $df/dJ_{STT}$ vs. $J_{SOT}$ under linear fitting. (g)-(j) PSD vs. SOT current density under fixed STT currents and an in-plane magnetic field $\mu_0 H_{ex} = -250$ mT in another sample. (k) Oscillating frequency vs. $J_{SOT}$ under different $J_{STT}$ (solid lines are theoretical calculation) and (l) the extracted $df/dJ_{SOT}$ vs. $J_{STT}$.

### 3.3 Electrical modulation of nonlinearity

The modulation of nonlinearity $N$ by the combination of STT and SOT current in 1.3 nm CoFeB is also investigated. Figure 4(a)-(d) shows the PSD vs. $J_{STT}$ under fixed SOT current. It is obvious that a positive (negative) value of SOT current can facilitate (suppress) the oscillation. When $J_{SOT} = 15$ MA/cm$^2$, particularly, the emission power improves and the threshold STT current density $J_{th,STT}$ of oscillation decreases significantly, which is consistent with our previous study [28]. The output frequency at more different values of $J_{SOT}$ is shown in Figure 4(e). The extracted $df/dJ_{STT}$ in Figure 4(f) decreases with the increase in $J_{SOT}$, demonstrating the tunability of SOT current on nonlinearity $N$. On the other hand, we further measured the PSD spectra when sweeping SOT current under fixed STT currents, as illustrated in Figure 4(g)-(j). As a result, the SOT current could also tune and exhibit a linear relationship with output frequency, which allows us to analyze the nonlinearity in the same manner as with STT. Similar to the performance when sweeping STT current, the change rate of frequency manifests a clear dependency on $J_{STT}$. As presented in Figure 4(k) and (l), the slope $df/dJ_{SOT}$ gradually decreases with the increase of $J_{STT}$, also indicating a continuous modulation of nonlinearity $N$. To explain this phenomenon, we reconsidered the combined effects of SOT and STT currents on $\zeta$ and revised eq. (1) as follows:

$$f(J_{STT}) = f_{FMR} + \frac{N}{2\pi}\left(1 - \frac{1+Q}{\frac{J_{STT}+\eta J_{SOT}}{J_{th,STT}} + Q}\right), \quad (5)$$

where $J_{STT}$ become the independent variable, and a correction term $\eta J_{SOT}$ is added to $\zeta$. Therefore, a positive (negative) SOT current could generate extra spin torque promoting (suppressing) the oscillation, which is equivalent to a change in the effective STT current density $\Delta J = \eta J_{SOT}$, and results in different intervals of the precession frequency. Theoretical calculation according to eq. (5) is shown as the solid lines in Figure 4(e), which are in good agreement with the experimental results (the small deviation could be attributed to thermal effect induced by both STT and SOT current). The proportional relationship between $\Delta J$ and $J_{SOT}$ shown in the inset also confirms this explanation. Besides, the mechanism of STT modulation is similar, and the theoretical prediction is also plotted as solid lines in Figure 4(k). Notice that $df/dJ_{STT}$ can no longer be simplified as a constant during the analytical calculation, but it can still be approximately estimated through linear fitting, and modulated by both SOT and STT current. Other explanations of this electrical modulation are also worth discussing, such as additional effective field $H_{\text{eff}}$ generates by field like torque and current-induced Joule heat. However, based on our previous study [41], the maximum variation in the $H_{\text{eff}}$ is only about 3 mT, which can be neglected compared to the strong $H_{\text{ex}}$. And the asymmetric impact of positive and negative currents on nonlinearity (Figure 4(f) and (l)) largely rules out the influence of thermal effects. As a result, we proved that both SOT and STT current could modulate the nonlinearity, and the proposed model aligns well with the experiments. Such dependency provides an electrical approach to modulate nonlinearity in three-terminal STNOs, which is more convenient and better for miniaturization than tuning by $H_{\text{ex}}$. These findings open up a new dimension for engineering three-terminal MTJ-STNOs, and could inspire more flexible designs for emerging STNO-based computing.

## 4 Conclusions

In conclusion, we have experimentally studied the modulation of nonlinearity $N$ in three-terminal MTJ-STNOs. In 1.3 nm CoFeB, the magnitude of magnetic field does not contribute to modify $N$, while the direction of magnetic field can continuously change $N$ from positive to negative. Besides, nonlinearity can also be tuned by the thickness of CoFeB free layer, and reaches zero in 1.1 nm CoFeB, providing us a more intrinsic and robust approach to attain zero nonlinearity in STNOs. Moreover, by combining the SOT and STT current, the nonlinearity could be efficiently modified by either STT or SOT current, and such phenomenon can be explained by a rectified model considering both currents. This electrical way of tuning the nonlinearity is more convenient, better for miniaturization, and could also inspire more intriguing designs for signal detection, spectrum analysis, energy harvesting and neuromorphic devices. Our findings provide a comprehensive understanding and open up a new dimension for the current tunability of nonlinearity in MTJ-STNOs, and will pave the way for further optimization in STNO-based applications.